\RequirePackage{fix-cm}
\documentclass[onecolumn,fleqn,usenatbib]{article} 

\RequirePackage{graphicx}

\usepackage{rotating}

\usepackage{xcolor}
\usepackage{newtxtext,newtxmath}

\usepackage{graphicx}
\usepackage{amsmath}
\usepackage{amssymb}	
\usepackage{rotating}

\title{Dynamic Dark Energy Equation of State (EoS) and Hubble Constant analysis using type Ia supernovae from Union 2.1 dataset.}
\date{2019\\ October}
\author{
	Syed Faisal ur Rahman$^{1}$\thanks{E-mail: faisalrahman36@hotmail.com}
	\\
	$^{1}$	Institute of Space and Planetary Astrophysics $(ISPA)$, \\University of Karachi $(UoK)$, Karachi, Pakistan\\
	Karachi Institute of Technology and Entrepreneurship $(KITE)$,Karachi, Pakistan	
}
\usepackage{multicol}
\usepackage{chngpage}

\begin{document}

\maketitle

\begin{abstract}
This paper constraints dynamic dark energy equation of state (EoS) parameters using the type Ia supernovae from Union 2.1 dataset. The paper also discusses the dependency of dynamic dark energy EoS parameters on the chosen or assumed value of the Hubble Constant. To understand the correlation between the Hubble Constant values and measured dynamic dark energy EoS parameters, we used recent surveys being done through various techniques such as cosmic microwave background studies, gravitational waves, baryonic acoustic oscillations and standard candles to set values for different Hubble Constant values as fixed parameters with CPL and WCDM models. Then we applied trust region reflective (TRF)  and dog leg (dogbox) algorithms to fit dark energy density parameter and dynamic dark energy EoS parameters. We found a significant negative correlation between the fixed Hubble Constant parameter and measured EoS parameter, w0. Then we used two best fit Hubble Constant values (70 and 69.18474) km $s^{-1}$ $Mpc^{-1}$ based on Chi-square test to test more dark energy EoS parameters like: JBP, BA, PADE-I, PADE-II, and LH4 models and compared the results with $\Lambda$-CDM with constant $w_{de}$=-1, WCDM and CPL models. We conclude that flat $\Lambda$-CDM and WCDM models clearly provide best results while using the BIC criteria as it severely penalizes the use of extra parameters. However, the dependency of EoS parameters on Hubble Constant value and the increasing tension in the measurement of Hubble Constant values using different techniques warrants further investigation into looking for optimal dynamic dark energy EoS models to optimally model the relation between the expansion rate and evolution of dark energy in our universe. 
\end{abstract}

\section{Introduction}

The discovery of the accelerated expansion of the universe \cite{Riess et al. 1998} \cite{Perlmutter 1999}\cite{PerlmutterSchmidt 2003} revolutionized modern cosmology and answered many questions related to the evolution of our universe. However, we are still trying to understand the ingredient which is likely responsible for the accelerated expansion of the universe i.e. dark energy. Dark energy seems to be something which is not only overcoming the tendency of collapse of the matter in our universe but it is also providing a push for the accelerated expansion of our universe \cite{Weinberg 2008}. After the discovery of the accelerated expansion of the universe in the late 1990s by HighZ Supernova and the Supernova Cosmology Project teams \cite{Riess et al. 1998}\cite{Perlmutter 1999} \cite{PerlmutterSchmidt 2003} using the type Ia supernovae, several observations applying various signatures like cosmic microwave background radiation (CMB), baryonic acoustic oscillation (BAO), Cepheid Variables, large scale structures etc. \cite{Bennett et al. 2013} \cite{Hinshaw et al. 2013}\cite{Planck 2018} \cite{Birrer et al. 2018}\cite{Macaulay et al. 2019}\cite{Riess et al. 2019}, confirmed the accelerated expansion of our universe. 
Although, these observations confirm that our universe is going through a phase of accelerated expansion but these different observations also presented some serious problems by getting variations in their measurements of the cosmological parameters based on the standard model of cosmology or the $\Lambda$-CDM model \cite{Liddle 2003}\cite{Jackson 2015}\cite{Rahman 2018} which is providing impetus towards the development of greater interest in non $Lambda$-CDM model studies \cite{Zhai et al. 2017}\cite{Khosravi et al. 2019}\cite{Sola et al. 2019}. New standard candles like active galactic nuclei (AGN) and the use of gravitational waves as standard sirens\cite{Abbott et al. 2016}\cite{LIGO 2017} are also being explored to get better measurements of cosmological parameters at high redshifts \cite{Watson et al. 2011}.

\section{Cosmology from type Ia Supernova}
Type Ia Supernovae are useful tools to be used as standard candles because of their almost standard absolute magnitude values. Therefore observations of apparent magnitude (m) and redshift (z) for type Ia Supernovae can lead to meausrements of key cosmological parameters: $\Omega$$\Lambda$, $\Omega$r,  and $\Omega$m, the dark energy, radiation and matter density parameters respectively within the Lambda-CDM cosmology framework \cite{Weinberg 2008}\cite{Perlmutter 1999} \cite{PerlmutterSchmidt 2003}\cite{Riess et al. 1998}.
The difference between apparent magnitude (m) and  absolute magnitude is known as the distance modulus, $µ$ \cite{Weinberg 2008} \cite{Davis 2012}:
\begin{equation}
\mu= m - M                 
\end{equation}							
Given a set of assumed cosmological parameters (C), the redshift of an object, its apparent magnitude and luminosity distance DL are linked thus:
\begin{equation}
m(C,z)=5log(DL(C,z))+M+25 
\label{m_Cz)}
\end{equation}
    						               			
Thus luminosity distance and distance modulus are linked:
\begin{equation}
\mu(C,z)= 5log[DL(C,z)] + 25
\label{mu_Cz}
\end{equation}
							         				
Here luminosity distance (DL) is in Mpc.For a spatially flat universe, we can write luminosity distance as:
\begin{equation}
DL(z) = (1+z)\chi(z)
\label{eq:DLCosmology}
\end{equation}		
Where, $$\chi(z)=c\eta(z)$$ is the comoving distance and $\eta(z)$ is conformal loop back time which can be calculated as:
\begin{equation}
\eta(z) = \int_{0}^{z} \frac {dz\textquoteright}{H(z\textquoteright)}
\end{equation}
      							         			
Here, $E(z)=\sqrt{\Omega\Lambda I(z)+\Omega r(1+z)^4+\Omega m(1+z)^3}$  for flat Lambda-CDM model. I(z) depends on the parametrization of the dark energy equation of state (EoS) and for standard $\Lambda$-CDM model with EoS as $w_{de}$(z)=-1 (constant), the multiplier I(z) becomes $‘1’$ .

We can separate contribution of H0 and absolute magnitude 'M' from the \eqref{m_Cz)} and \eqref{mu_Cz}, as \cite{Perlmutter 1999} \cite{Davis 2012}:

\begin{equation}
\bar{\textbf{\textit{M}}}=M+25+5log(c/H0)+\sigma_{M} 
\label{eq:sigma_M}
\end{equation}
Here 'c' is the speed of light in vacuum and $\sigma_{M}$ is the uncertainty in  absolute magnitude of type Ia supernovae. This is often done to marginalize uncertainties arising from measurements of H0 and M. However, the dominant contributor in these uncertainties is H0. We are fixing different H0s from various surveys to test
them for most suitable H0 for our dataset in relation with the equation of state (EoS) models in discussion which will minimize these uncertainties for the most suitable value of H0. Therefore instead of separating $\bar{\textbf{\textit{M}}}$, we can fit cosmologies using \eqref{m_Cz)}and \eqref{mu_Cz}. The contribution from absolute magnitude uncertainties is very minor if we apply proper fits for coefficients for stretch, color and probability of supernova in data are hosted by galaxies with less than certain threshold mass. We use Union 2.1's compilation \cite{Suzuki et al. 2012} magnitude vs redshift table which used fitted values for coefficients of stretch, color and the probability that a particular supernova in dataset was hosted by a low-mass galaxy. The dataset also employs a constant M$\approx$-19.31 with  uncertainties in distance modulus arising from fitting values and systematic contributions mentioned separately as distance modulus error which we incorporated in our model fitting using TRF and dogbox \cite{VoglisLagaris 2004} \cite{Branch 1999} \cite{Nocedal 2006} and $\chi^2$ analysis, and so it is absorbed in the parameter error bounds provided H0 is set to an optimal value.
\section{Dataset and Data Analysis Techniques}

For our study, we use Union 2.1 \cite{Suzuki et al. 2012} dataset which is publicly shared by Supernova Cosmology Project (SCP)\cite{Perlmutter 1999} \cite{PerlmutterSchmidt 2003}\cite{Amanullah et al. 2010}. The dataset is comprised of 580 type Ia supernovae which passed the usability cuts. The dataset is comprised of redshift range $0.015\leq z \leq 1.414$ with median redshift at $z\approx 0.294$.
 
We use SciPy's \cite{Jones et al. 2001} optimize package's trust region reflective (TRF) and dog leg (dogbox) algorithms \cite{VoglisLagaris 2004} \cite{Branch 1999} \cite{Nocedal 2006}, which are suitable for problems with constraints as in our case, to fit dark energy density parameter and dynamic dark energy EoS parameters for $\Lambda$-CDM, WCDM, CPL,JBP,BA,PADE-I,PADE-II and LH4 models \cite{BarbozaAlcaniz 2008}\cite{ChevallierPolarski 2001}\cite{Linder 2003} \cite{Jassal et al. 2005a}\cite{Jassal et al. 2005b} \cite{LinderHuterer 2005} \cite{Wei et al. 2014}. We also apply grid method to obtain log likelihood \cite{DavisParkinson 2016} for WCDM and CPL to compare results obtained through TRF and dog leg methods \cite{VoglisLagaris 2004}. We used TRF and dogbox options simultaneously with our selected models and then used the best fit results based on the $\chi^{2}$ values. TRF is an iterative algorithm which based on the first order optimality condition in a bounds constrained non-linear minimization problem which forms a trust region shape based on the direction of the gradient and the distance from the bounds. TRF considers and ellipsoid trst region for bounds. The other variant dogbox (dog-leg) considers a rectangular region for bounds \cite{VoglisLagaris 2004} \cite{Branch 1999} \cite{Nocedal 2006}. We applied both TRF and dogbox configurations and selected the optimal set of parameters constraints based on $\chi^{2}$ values for Lambda-CDM,WCDM and CPL methods. Then applied TRF on dynamic dark energy EoS models with more variables as based on our initial test with lesser variable models, TRF is found more robust of the two and almost gives similar results. TRF and dog-leg algorithms are not commonly used in constraining cosmological parameters but we found these methods in general agreement with log likelihood method which were used for comparison purposes.

\section{Dynamic Dark Energy Equation of State (EoS)}
In order to extend the standard Lambda-CDM model to incorporate dynamic dark energy EoS, we can define I(z) as \cite{Zhai et al. 2017}\cite{Heacox 2015}:

\begin{equation}
I(z)=exp(3\int_{0}^{z} \frac{1+w_{de}(z\textquoteright)}{1+z\textquoteright} dz\textquoteright)
\end{equation}

For the study we tested various dynamic dark energy EoS models.

We started with standard flat Lambda-CDM model with $w_{de}$=-1 and then tested WCDM model by treating $w_{de}$ as free parameter. Then we moved towards more complex CPL, JBP, BA,PADE-I,PADE-II and LH4  models \cite{BarbozaAlcaniz 2008}\cite{ChevallierPolarski 2001}\cite{Linder 2003} \cite{Jassal et al. 2005a}\cite{Jassal et al. 2005b} \cite{LinderHuterer 2005} \cite{Wei et al. 2014} with model equations as:

CPL \cite{ChevallierPolarski 2001} \cite{Linder 2003}
\begin{equation}
w_{de}(z)=w_{0} + w_{a}\frac{z}{(1+z)}
\end{equation}	

JBP \cite{Jassal et al. 2005a}\cite{Jassal et al. 2005b}

\begin{equation}
w_{de}(z)=w_{0} + w_{a}\frac{z}{(1+z)^2}
\end{equation}

BA \cite{BarbozaAlcaniz 2008}
\begin{equation}
w_{de}(z)=w_{0} + w_{a}\frac{z(1+z)}{(1+z^2)}
\end{equation}

PADE-I \cite{Wei et al. 2014}

\begin{equation}
w_{de}(z)= \frac{w_{0}+w_{a}\frac{z}{(1+z)}}{1.+w_{b}\frac{z}{(1+z)}}
\end{equation}

For wb=0, PADE-I reduces to CPL model.

\bigskip 

PADE-II \cite{Wei et al. 2014}

\begin{equation}
w_{de}(z)= \frac{w_{0}+w_{a}ln(\frac{1}{1+z})}{1.+w_{b}ln(\frac{1}{(1+z)})}
\end{equation}

Linder-Huterer (LH4) \cite{LinderHuterer 2005}

\begin{equation}
w_{de}(z)=w_{0} + \frac{ (w_{a}-w_{0})} {1+\frac{1}{(1+z)a_{t}}^{1/T}}
\end{equation}


For parameter boundaries for TRF and dog leg analysis, we set $\Omega \Lambda$ boundary between 0.65 and 0.75. For w0, we set the upper boundary as w0<-1/3 which is a pre-condition for accelerated expansion of our universe but for lower limits we first set restrict it to $w0 \geq -1$ to exclude phantom dark energy \cite{Vikman 2005}\cite{Farnes 2018} and keeping it in  quintessence regime \cite{Weinberg 2008}. Then we set as $\infty <w0 \leq -1/3$ to allow phantom dark energy. This was done to minimize boundary condition bias while running the optimization algorithms. Similarly for wa, we chose two set of boundaries $-5 \leq wa \leq 5$ and $-0.3 \leq wa \leq 0.3$ to avoid localization bias for optimization algorithm. In case of PADE I and II, wb boundaries are set as $-1< wb <0$ while others remain same. In LH4 case, we set both T and $a_{t}$ between and 0 and 1.

\section{Hubble Constant Value}

\begin{figure}
	\centering
	\includegraphics[width=1\linewidth]{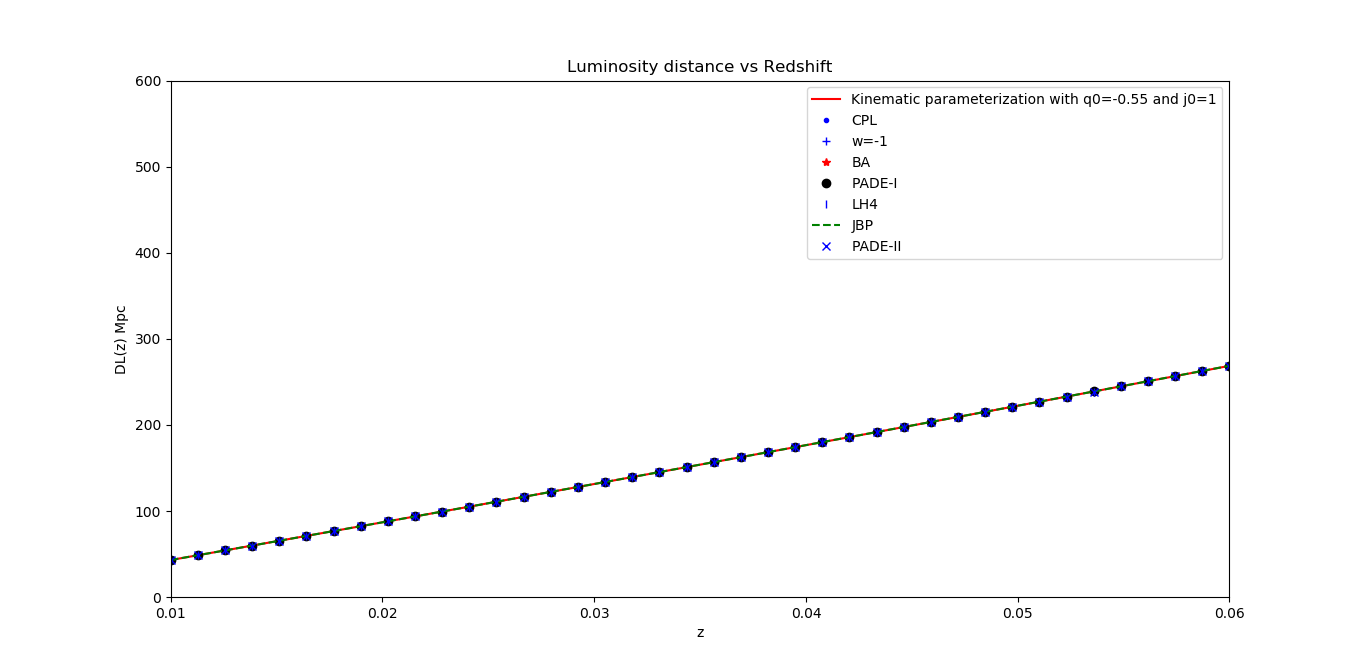}
	\caption{Luminosity distance plots using kinematic expression for DL(z) and comparison with DL(z) using various EoS models and $Lambda$-CDM with with $w_{de}$(z) as a constant value of w=-1.}
	\label{fig:luminosityvsredshiftzmax0-06}
\end{figure}

The value of Hubble Constant has recently been  a topic of great interest in physics and astronomy community. It had been studied in the past like the first precise measurements by Sandage 1958 \cite{Sandage 1958} which gave H0=75 but recent interest has increased as the measurements of H0 from cosmic microwave background (CMB), baryon acoustic oscillations (BAO),  standard candles and others do not seem to agree with each other \cite{Freedman 2017}\cite{Jackson 2015} \cite{Planck 2018} \cite{WojtakAdriano 2019}\cite{Vattis et al. 2019}\cite{Riess et al. 2019}.The problem has become even more interesting as the expansion rate is found to be same in all directions by \cite{Soltis et al. 2019} based on 1000 type Ia supernovae samples. Therefore we considered it appropriate to measure CPL and WCDM model parameters by fixing H0 values from Planck 2018, Riess 2018,Abbott et al. 2017,Planck+SNe+BAO-Planck 2018, Planck+BAO/RSD+WL-Planck 2018, H0LiCOW 2018 and DES 2018 \cite{Abbott et al. 2017} \cite{Birrer et al. 2018}. We also fit our own value for Union 2.1 dataset \cite{Suzuki et al. 2012} using the kinematic expression from Riess et al. 2016 \cite{Riess et al. 2016} for luminosity distance with source redshift of z<0.04. Figure \ref{fig:luminosityvsredshiftzmax0-06} shows that luminosity distances from \eqref{eq:DLKinematic} is in good agreement with luminosity distances from \eqref{eq:DLCosmology} for z<0.04 using various EoS models.

The kinematic expression from \cite{Weinberg 2008} \cite{Riess et al. 2016} is written as:

\begin{equation}
DL(z)=\frac{cz}{H0}[1+\frac{(1-q_{0})z}{2}-\frac{(1 - q_{0} - 3q_{0}^{2}+ j_{0})z^{2}}{6}+O(z^{3})]
\label{eq:DLKinematic}
\end{equation}

With $q_{0} = -0.55$ and $j_{0} = 1$.

We can see from tables \ref{tab:Table1} and \ref{tab:Table2} that our best measurements based on $\chi^{2}$ values for both CPL and WCDM are obtained through H0=70 km $s^{-1}$ $Mpc^{-1}$ which is measured by Abbott et al. 2017 by studying gravitational waves (GW170817) \cite{LIGO 2017}\cite{Abbott et al. 2016}\cite{Abbott et al. 2017} from neutron stars collision and was also measured by  Wilkinson Microwave Anisotropy Probe (WMAP) \cite{Bennett et al. 2013} with WMAP only dataset. Our second best measurements were obtained through the best fit $H0=69.18473827 ± 0.50179901$ or approximately 69.185 km $s^{-1}$ $Mpc^{-1}$ value from Union 2.1 dataset using kinematic expression for luminosity distance which is closer to the value obtained by \cite{Bennett et al. 2013} using WMAP+eCMB+BAO+H0 data set \cite{Hinshaw et al. 2013}. We applied TRF with bounds $65 \leq H0 \leq 75$ to obtain the best fit H0 value. Both of these values are interestingly somewhat in the middle region of the H0 values obtained by early universe studies \cite{GorbunovRubakov 2011} like Planck cosmic microwave background (CMB) \cite{Planck 2014a}\cite{Planck 2014b}\cite{Planck 2016}\cite{Planck 2018} or baryon acoustic oscillations (BAO) \cite{Grieb et al. 2017},\cite{Macaulay et al. 2019} which give H0$\approx$ 67  and standard candles studies like \cite{Riess et al. 1998} \cite{Riess et al. 2007} \cite{Riess et al. 2016} \cite{Riess et al. 2018a}\cite{Riess et al. 2018b} \cite{Pietrzynski et al.2019}\cite{Riess et al. 2019} which give H0 $>$73. Because of this discrepancy in the measurement of H0,higher redshift studies of type Ia supernovae and other standard candles are becoming important \cite{RisalitiLusso 2019}\cite{Riess et al. 2018a}\cite{Daniel et al. 2019}. Like early universe studies, standard candles are also useful to study the nature of dark energy \cite{Wood-Vasey et al. 2007} which is still an open problem of cosmology \cite{Davis et al. 2007}\cite{DavisParkinson 2016}. 

\begin{table}
	\centering
	\caption{Best fit values for WCDM model using union 2.1 dataset}
	\label{tab:Table1}%
	\begin{tabular}{lllll}
		\hline
		WCDM &         &         &         &  \\
		\hline
		H0 & $\Omega \Lambda$ & w0 & $\chi^2$ & Bounds on $\Omega \Lambda$,w0 \\
		\hline
		67.400  & 0.75    & -0.7024 & 606.6761 & (0.65,0.75),(-$\infty$,-1/3) \\
		67.400  & 0.75    & -0.7024 & 606.6761 & (0.65,0.75),(-1,-1/3) \\
		73.520  & 0.65    & -1.7459 & 614.5908 & (0.65,0.75),(-$\infty$,-1/3) \\
		73.520  & 0.75    & -1.0000 & 740.6163 & (0.65,0.75),(-1,-1/3) \\
		\textbf{70.000} & \textbf{0.72} & \textbf{-1.0045} & \textbf{562.2257} & \textbf{(0.65,0.75),(-$\infty$,-1/3)} \\
		70.000  & 0.72    & -1.0000 & 562.2267 & (0.65,0.75),(-1,-1/3) \\
		72.500  & 0.65    & -1.5683 & 588.7033 & (0.65,0.75),(-$\infty$,-1/3) \\
		72.500  & 0.75    & -1.0000 & 646.1989 & (0.65,0.75),(-1,-1/3) \\
		67.770  & 0.75    & -0.7353 & 594.2718 & (0.65,0.75),(-$\infty$,-1/3) \\
		67.770  & 0.75    & -0.7353 & 594.2718 & (0.65,0.75),(-1,-1/3) \\
		\textbf{69.185} & \textbf{0.75} & \textbf{-0.8645} & \textbf{565.9402} & \textbf{(0.65,0.75),(-$\infty$,-1/3)} \\
		\textbf{69.185} & \textbf{0.75} & \textbf{-0.8645} & \textbf{565.9402} & \textbf{(0.65,0.75),(-1,-1/3)} \\
	\end{tabular}%

\end{table}%
\begin{table}
	
	\scriptsize	
	\centering
	\caption{Best fit values for CPL model using union 2.1 dataset}
	\label{tab:Table2}
	\begin{tabular}{lllllllll}

		\hline
		CPL  \\
		\hline
		H0 & $\Omega \Lambda$ & w0 & wa & $\chi^2$ & Bounds on $\Omega \Lambda$,w0,wa &         &  \\
		\hline

	66.300  & 0.65    & -0.333  & -3.601  & \textbf{620.9512} & (0.65,0.75),(-$\infty$,-1/3),(-5,5) \\

	66.300  & 0.65    & -0.333  & -3.601  & \textbf{620.9512} & (0.65,0.75),(-1,-1/3),(-5,5) \\
	
	66.300  & 0.75    & -0.567  & -0.300  & \textbf{650.2256} & (0.65,0.75),(-$\infty$,-1/3),(-0.3,0.3) \\
	
	66.300  & 0.75    & -0.567  & -0.300  & \textbf{650.2256} & (0.65,0.75),(-1,-1/3),(-0.3,0.3) \\
	
	67.400  & 0.65    & -0.333  & -4.678  & \textbf{585.7344} & (0.65,0.75),(-$\infty$,-1/3),(-5,5) \\
	
	67.400  & 0.65    & -0.333  & -4.678  & \textbf{585.7344} & (0.65,0.75),(-1,-1/3),(-5,5) \\
	
	67.400  & 0.75    & -0.664  & -0.300  & \textbf{602.6112} & (0.65,0.75),(-$\infty$,-1/3),(-0.3,0.3) \\
	
	67.400  & 0.75    & -0.664  & -0.300  & \textbf{602.6112} & (0.65,0.75),(-1,-1/3),(-0.3,0.3) \\
	
	67.770  & 0.65    & -0.419  & -4.326  & \textbf{579.2176} & (0.65,0.75),(-$\infty$,-1/3),(-5,5) \\
	
	67.770  & 0.65    & -0.419  & -4.326  & \textbf{579.2176} & (0.65,0.75),(-1,-1/3),(-5,5) \\
	
	67.770  & 0.75    & -0.697  & -0.300  & \textbf{590.9692} & (0.65,0.75),(-$\infty$,-1/3),(-0.3,0.3) \\
	
	67.770  & 0.75    & -0.697  & -0.300  & \textbf{590.9692} & (0.65,0.75),(-1,-1/3),(-0.3,0.3) \\
	
	68.340  & 0.65    & -0.584  & -3.491  & \textbf{571.3333} & (0.65,0.75),(-$\infty$,-1/3),(-5,5) \\
	
	68.340  & 0.65    & -0.584  & -3.491  & \textbf{571.3333} & (0.65,0.75),(-1,-1/3),(-5,5) \\
	
	68.340  & 0.75    & -0.749  & -0.300  & \textbf{577.1210} & (0.65,0.75),(-$\infty$,-1/3),(-0.3,0.3) \\
	
	68.340  & 0.75    & -0.749  & -0.300  & \textbf{577.1210} & (0.65,0.75),(-1,-1/3),(-0.3,0.3) \\
	
	\textbf{69.185} & \textbf{0.65} & \textbf{-0.830} & \textbf{-2.278} & \textbf{564.2394} & \textbf{(0.65,0.75),(-$\infty$,-1/3),(-5,5)} \\

	\textbf{69.185} & \textbf{0.65} & \textbf{-0.830} & \textbf{-2.278} & \textbf{564.2394} & \textbf{(0.65,0.75),(-1,-1/3),(-5,5)} \\

	69.185  & 0.75    & -0.827  & -0.300  & \textbf{565.2861} & (0.65,0.75),(-$\infty$,-1/3),(-0.3,0.3) \\

	69.185  & 0.75    & -0.827  & -0.300  & \textbf{565.2861} & (0.65,0.75),(-1,-1/3),(-0.3,0.3) \\
	
	\textbf{70.000} & \textbf{0.72} & \textbf{-1.005} & \textbf{-0.011} & \textbf{562.2257} & \textbf{(0.65,0.75),(-$\infty$,-1/3),(-5,5)} \\
	
	\textbf{70.000} & \textbf{0.72} & \textbf{-1.005} & \textbf{-0.011} & \textbf{562.2257} & \textbf{(0.65,0.75),(-$\infty$,-1/3),(-0.3,0.3)} \\
	
	70.000  & 0.72    & -1.000  & 0.039   & \textbf{562.2260} & (0.65,0.75),(-1,-1/3),(-5,5) \\
	
	70.000  & 0.72    & -1.000  & 0.039   & \textbf{562.2260} & (0.65,0.75),(-1,-1/3),(-0.3,0.3) \\

	72.500  & 0.67    & -1.727  & 2.402   & \textbf{583.9811} & (0.65,0.75),(-$\infty$,-1/3),(-5,5) \\

	72.500  & 0.65    & -1.598  & 0.300   & \textbf{587.6217} & (0.65,0.75),(-$\infty$,-1/3),(-0.3,0.3) \\

	72.500  & 0.75    & -1.000  & -1.142  & \textbf{623.7469} & (0.65,0.75),(-1,-1/3),(-5,5) \\

	72.500  & 0.75    & -1.000  & -0.300  & \textbf{635.4969} & (0.65,0.75),(-1,-1/3),(-0.3,0.3) \\
	
	73.520  & 0.65    & -2.101  & 3.575   & \textbf{603.5328} & (0.65,0.75),(-$\infty$,-1/3),(-5,5) \\
	
	73.520  & 0.65    & -1.773  & 0.300   & \textbf{613.0368} & (0.65,0.75),(-$\infty$,-1/3),(-0.3,0.3) \\
	
	73.520  & 0.75    & -1.000  & -1.950  & \textbf{682.8480} & (0.65,0.75),(-1,-1/3),(-5,5) \\

	73.520  & 0.75    & -1.000  & -0.300  & \textbf{722.6496} & (0.65,0.75),(-1,-1/3),(-0.3,0.3) \\
	\end{tabular}%

\end{table}%

\begin{figure}
	\centering
	\includegraphics[width=1\linewidth]{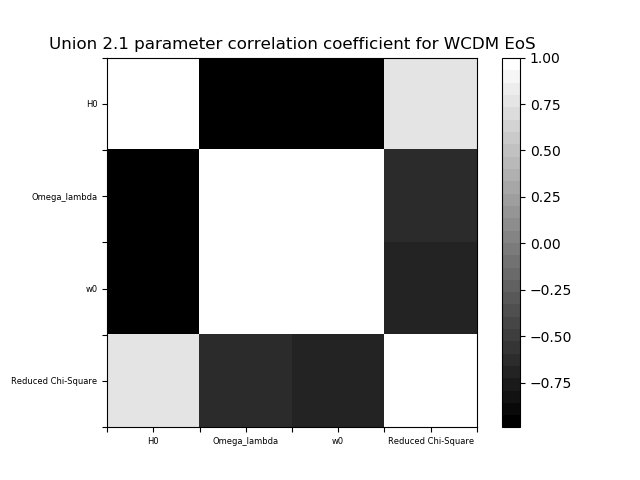}
	\caption{Cross-correlation of WCDM model parameters with H0, $\chi^{2}$ and each other. We can clearly observe significant negative cross-correlation between w0 and H0. }
	\label{fig:wcdmparameterscorrcoeff}
\end{figure}

\begin{figure}
	\centering
	\includegraphics[width=1\linewidth]{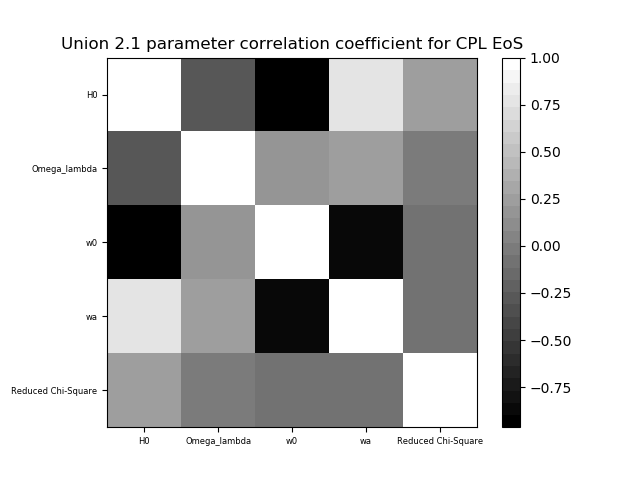}
	\caption{Cross-correlation of CPL model parameters with H0, $\chi^{2}$ and each other. We can again clearly observe significant negative cross-correlation between w0 and H0. We can also observe positive cross-correlation between H0 and wa.}
	\label{fig:cplparameterscorrcoeff}
\end{figure}

In order to understand how H0 value affects the measurements of dynamic dark energy EoS model parameters, we simply cross-correlated the data in tables \ref{tab:Table1} and \ref{tab:Table2}. Figures \ref{fig:wcdmparameterscorrcoeff} and \ref{fig:cplparameterscorrcoeff} show how the measurement or choice of the Hubble Constant can affect the measurements of dynamic dark energy EoS parameters in WCDM and CPL models. We can clearly observe significant negative cross-correlation between w0 and H0 for both WCDM and CPL models.


These results are particularly interesting due to the Hubble Constant tension arising due to the differences in measurements of H0 through cosmic microwave background, standard candles and other techniques. 

\section{Results}

\begin{figure}
	\centering
	\includegraphics[width=1\linewidth]{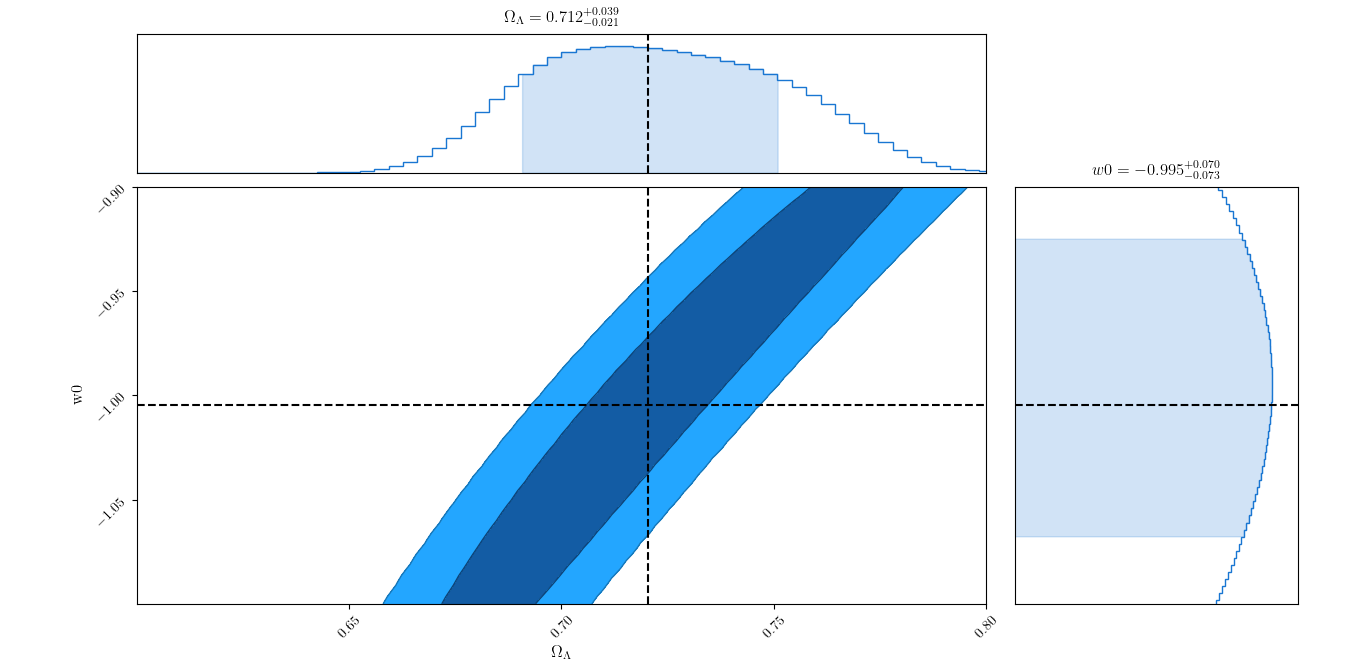}
	\caption{WCDM parameter constraints obtained through maximum likelihood and comparison with results from TRF and dog leg methods (dark dashed lines).}
	\label{fig:wcdmchainconsumerwithbestfit}
\end{figure}

In figure \ref{fig:wcdmchainconsumerwithbestfit}, for WCDM model, the maximum likelihood fit values are $\Omega \Lambda=0.712_{-0.021}^{+0.039}$ and $w0=-0.995_{-0.073}^{+0.070}$. Their corresponding mean likelihood fit values are $\Omega \Lambda$=0.724 ± 0.030 and w0=-1 ± 0.065. Both maximum and mean likelihood values agree, within one sigma overlapping values, with the best fit values obtained by using TRF and dog leg methods for H0=70. The best values from tables \ref{tab:Table1} and \ref{tab:Table3} are $\Omega \Lambda$=0.720362 ± 0.0626 and	w0=-1.00449 ± 0.1435. Values in table \ref{tab:Table1} are rounded off to fit in the columns.

\begin{figure}
	\centering
	\includegraphics[width=1\linewidth]{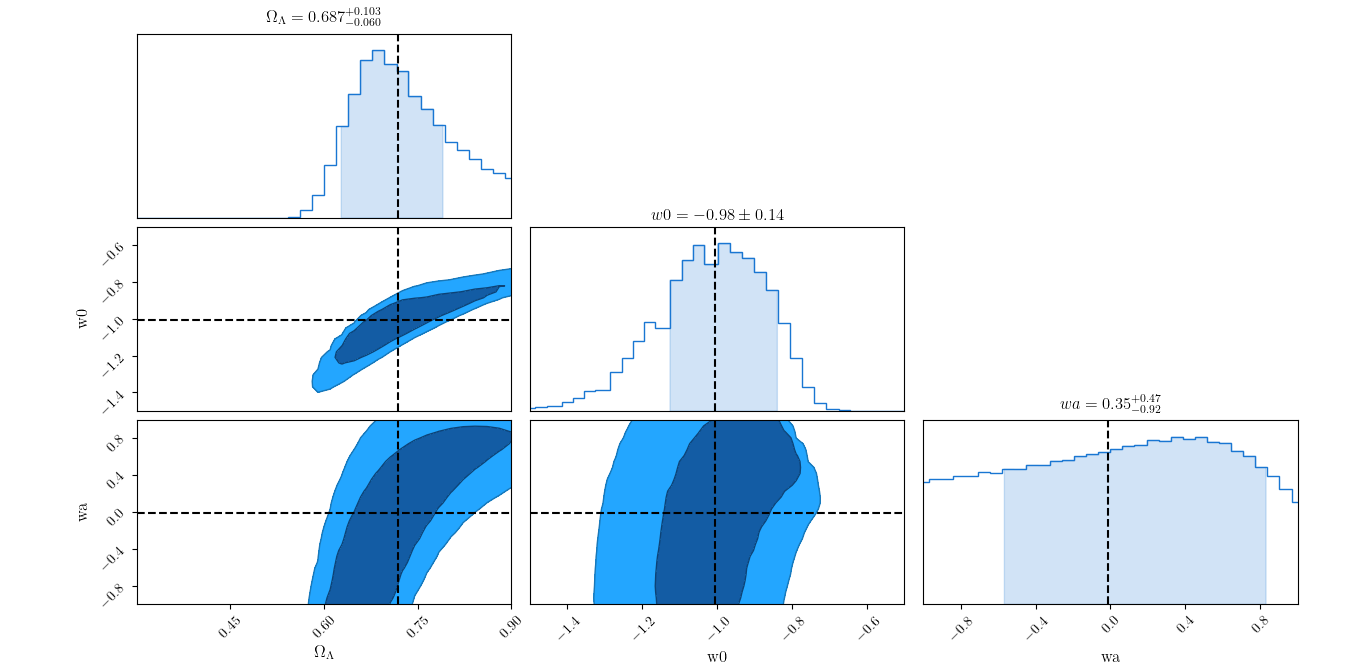}
	\caption{CPL parameter constraints obtained through maximum likelihood and comparison with results from TRF and dog leg methods (dark dashed lines).}
	\label{fig:cplchainconsumerwithbestfit}
\end{figure}

In figure \ref{fig:cplchainconsumerwithbestfit}, for CPL model, the maximum likelihood fit values are $\Omega \Lambda=0.687_{-0.060}^{+0.103}$, $w0=-0.98_{-0.014}^{+0.014}$ and $wa=-0.35_{-0.92}^{+0.47}$. Their corresponding mean likelihood fit values are $\Omega \Lambda$=0.731 ± 0.080, w0=-1.02 ± 0.015 and wa=0.01 ± 0.65. Again both maximum and mean likelihood values agree, within one sigma overlapping values, with the best fit values obtained by using TRF and dog leg methods for H0=70. The best values from tables \ref{tab:Table2} and \ref{tab:Table3} are $\Omega \Lambda$=0.71933 ± 0.27885,	w0=-1.00547 ± 0.291303 and wa= -0.01126 ± 3.033239. Values in table \ref{tab:Table1} are rounded off to fit in the columns. For wa, there is a relatively larger standard deviation in both likelihood estimates and in TRF and dog leg optimization approaches which is likely due to smaller redshift coverage from type Ia supernovae sample from Union 2.1. On very large redshifts, wa almost plays an equal role as w0 in CPL model because on extremely large 'z' values, $w_de$(z) approximately becomes $~$ w0 + wa. However in case of a model like JBP, the model will be more or entirely dependent on w0. This means higher redshift surveys especially highly sensitive all sky surveys like galaxy surveys to study the late time integrated Sachs-Wolfe effect (ISW) \cite{SachsWolfe 1967}\cite{Afshordi 2004}\cite{RahmanIqbal 2019} or surveys studying the early universe signatures like cosmic microwave background radiation (CMB) or baryonic acoustic oscillations (BAO), can play an important part in estimating parameters like wa or other extended EoS model parameters can make major contributions in higher redshifts in various dynamic dark energy equation of state (EoS) models which are in discussion in this study.

To see how dynamic dark energy EoS evolves in JBP, BA, PADE-I, PADE-II, and LH4 models especially in comparison the results from the flat $\Lambda$-CDM model with constant $w_{de}$=-1, WCDM and CPL models, we again applied TRF and dog leg methods \cite{VoglisLagaris 2004} simultaneously and selected the best fit values based on $\chi^{2}$ criteria. 

\begin{figure}
	\centering
	\includegraphics[width=1\linewidth]{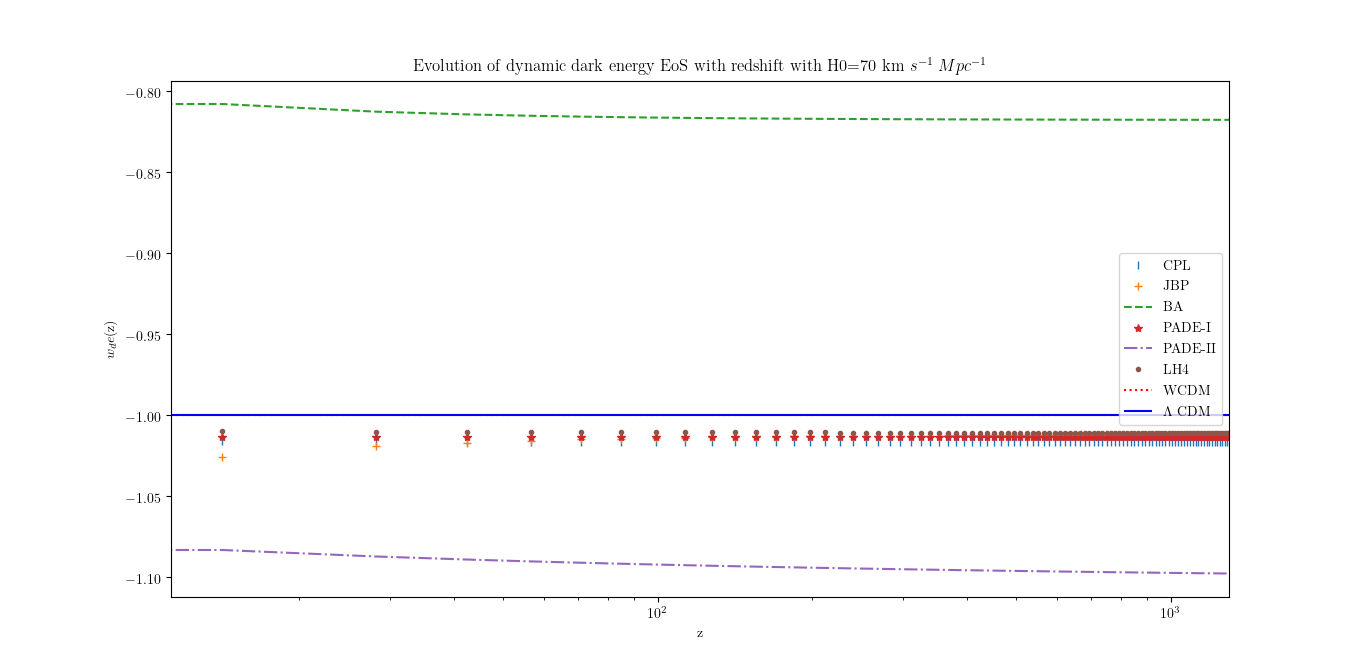}
	\caption{Evolution of $w_{de}$(z) for various dynamic dark energy EoS models with redshift  for H0=70 km $s^{-1}$ $Mpc^{-1}$.  }
	\label{fig:wdeh070}
\end{figure}

We can see in figure \ref{fig:wdeh070} that for H0=70 km $s^{-1}$ $Mpc^{-1}$, the results are closer to $\Lambda$-CDM model with constant $w_{de}$=-1 except for BA model which is in quintessence regime and PADE-II which is a bit farther than $w_{de}$=-1 in comparison with others. However, due to large standard deviations from mean for wa, wb, $a_{t}$ and T parameters in CPL, JBP,BA, PADE-I, PADE-II and LH4 models \cite{BarbozaAlcaniz 2008}\cite{ChevallierPolarski 2001}\cite{Linder 2003} \cite{Jassal et al. 2005a}\cite{Jassal et al. 2005b} \cite{LinderHuterer 2005} \cite{Wei et al. 2014} for relatively smaller redshift objects like in Union 2.1 dataset of type Ia supernovae, we still need to test these models using early universe signatures like CMB and BAO. For our type Ia supernova dataset with relatively smaller redshift coverage in comparison with they early universe studies, we can see that $\Lambda$-CDM model with $w_{de}$=-1 as fixed value is still the preferred model based on Bayesian information criterion (BIC) \cite{Schwarz 1978}\cite{Arevalo et al. 2017}\cite{Liddle 2007} especially if we consider $\Delta$BIC values which are basically the difference of BIC values from our models in discussion with the lowest BIC obtained from these models. $\Delta$BIC $>$ 2 suggests positive evidence against a model with higher BIC and $\Delta$BIC $>$ 6 suggests strong evidence against higher BIC value models \cite{KassRaftery 1995} as BIC heavily penalizes the inclusion of newer parameters \cite{Liddle 2007} despite having better $\chi^2$ scores for non $\Lambda$-CDM models. This can change for higher redshift or early universe studies when extra parameters in dynamic dark energy EoS models are potentially going to play important role which will also be useful for H0 studies \cite{GorbunovRubakov 2011} \cite{Planck 2018}\cite{Riess et al. 2019} \cite{RisalitiLusso 2019}\cite{Poulin et al. 2019}\cite{Liu et al. 2019} .   
\begin{figure}
	\centering
	\includegraphics[width=1\linewidth]{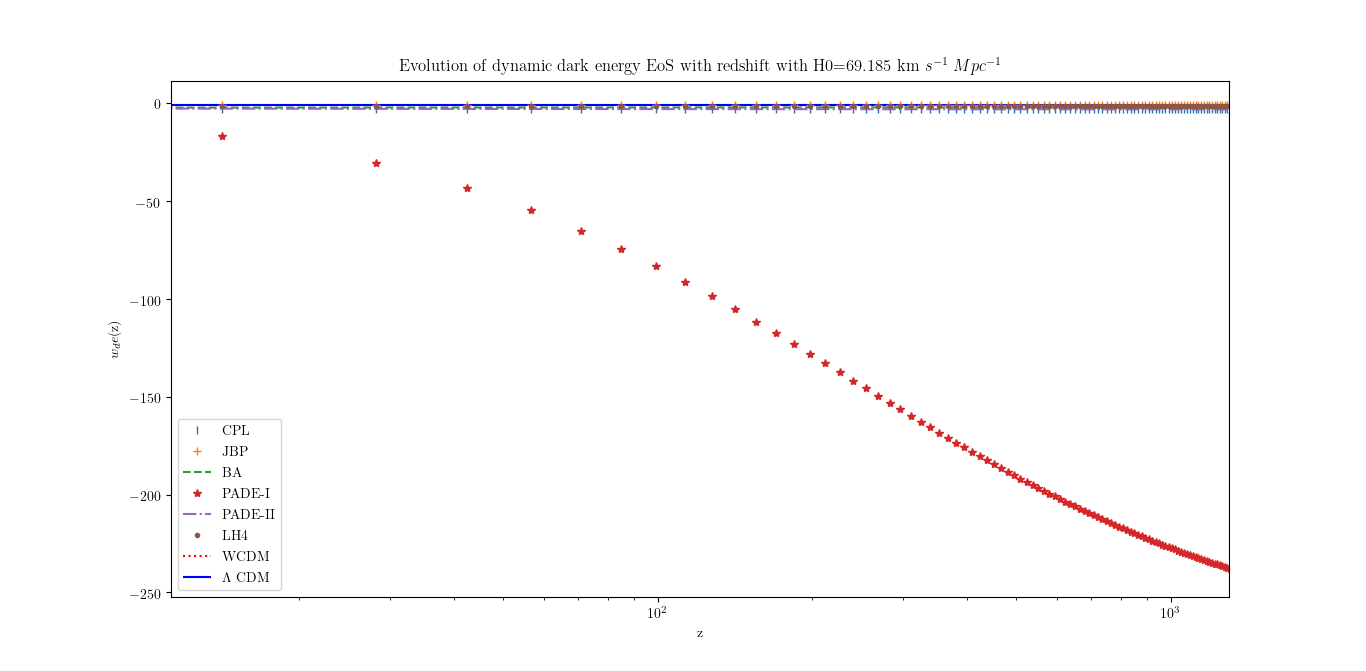}
	\caption{Evolution of $w_{de}$(z) for various dynamic dark energy EoS models with redshift  for H0=69.185 km $s^{-1}$ $Mpc^{-1}$.}
	\label{fig:wdeh069-185}
\end{figure}
\begin{figure}
	\centering
	\includegraphics[width=1\linewidth]{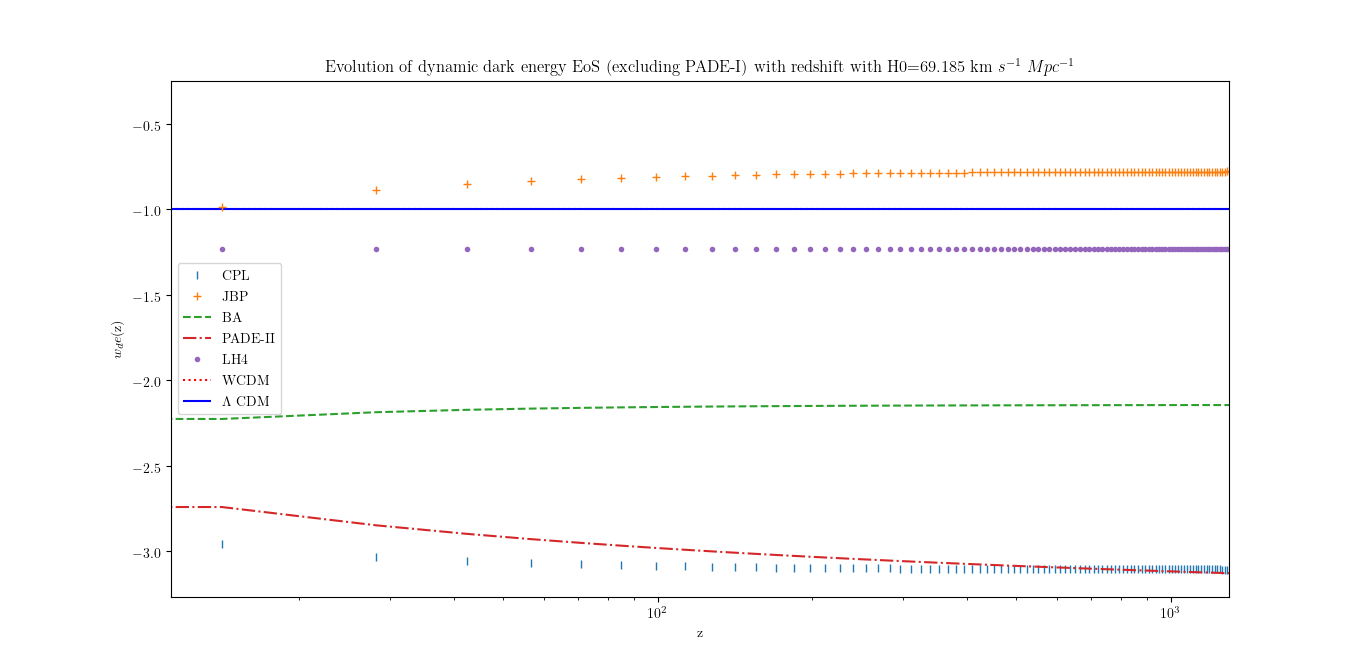}
	\caption{Evolution of $w_{de}$(z) for various dynamic dark energy EoS models with redshift  for H0=69.185 km $s^{-1}$ $Mpc^{-1}$ excluding PADE-I.}
	\label{fig:wdeh069-185nopade-i}
\end{figure}

For H0=69.185 km $s^{-1}$ $Mpc^{-1}$, we first look at figure \ref{fig:wdeh069-185} and observe that PADE-I is showing most deviation from $w_{de}$=-1 in comparison with the others especially at higher redshifts. This difference in scale of deviation towards $w_{de}$<<-1 is due to the relatively higher contribution of wa and wb of PADE-I model with increasing redshift values. In figure \ref{fig:wdeh069-185nopade-i}, we remove PADE-I model to see the evolution of $w_{de}$(z) in other models. We can see that apart from JBP, which is moving towards quintessence regime, others are closer to phantom regime \cite{Vikman 2005}\cite{Farnes 2018} with BA and PADE-II deviating away more from $w_{de}$=-1 and towards phantom regime. Theoretically,all structures in our universe would be eventually ripped apart by the repulsive forced associated with the phantom dark energy \cite{Vikman 2005}\cite{Weinberg 2008}. It will be interesting to see if future high precision standard candles, early universe and other surveys can settle expansion rate debate and which $w_{de}$ evolution or best fit value will be associated with it as we can observe from figures \ref{fig:wcdmparameterscorrcoeff} and \ref{fig:cplparameterscorrcoeff} that expansion rate and dark energy EoS parameters have significant cross-correlation with each other.  

We can also see from figures \ref{fig:wdeh070}, \ref{fig:wdeh069-185} and \ref{fig:wdeh069-185nopade-i} that despite H0 values being $<2\% $ different from each other, their impact on $w_{de}$(z) evolution is significant for all the models. This difference is significant enough to impact our understanding of the scales and evolution of our universe which warrants the need to carefully model $w_{de}$(z) in observations of  early universe signatures, galaxy surveys, standard candles, standard rulers and recently discovered gravitational waves which can be used as standard sirens \cite{Schutz 1999} \cite{Jarvis et al. 2014} \cite{Rahman 2018}\cite{Chen et al. 2018}. Gravitational waves can also be used to study the gravitational wave strain signals from type Ia supernovae and we can  use them to study cosmological parameters. For this purpose it will be useful to carefully study the progenitors of the type Ia supernovae (\cite{KeiichiTerada 2016}\cite{Rahman 2018}) as the mass profiles of the objects involved will be crucial in modeling the expected signal \cite{Schutz 1999}\cite{Rahman 2018}. 

\section{Effect of degeneracy between H0 and M on best fit values}

\begin{figure}
 	\centering
 	\includegraphics[width=0.7\linewidth]{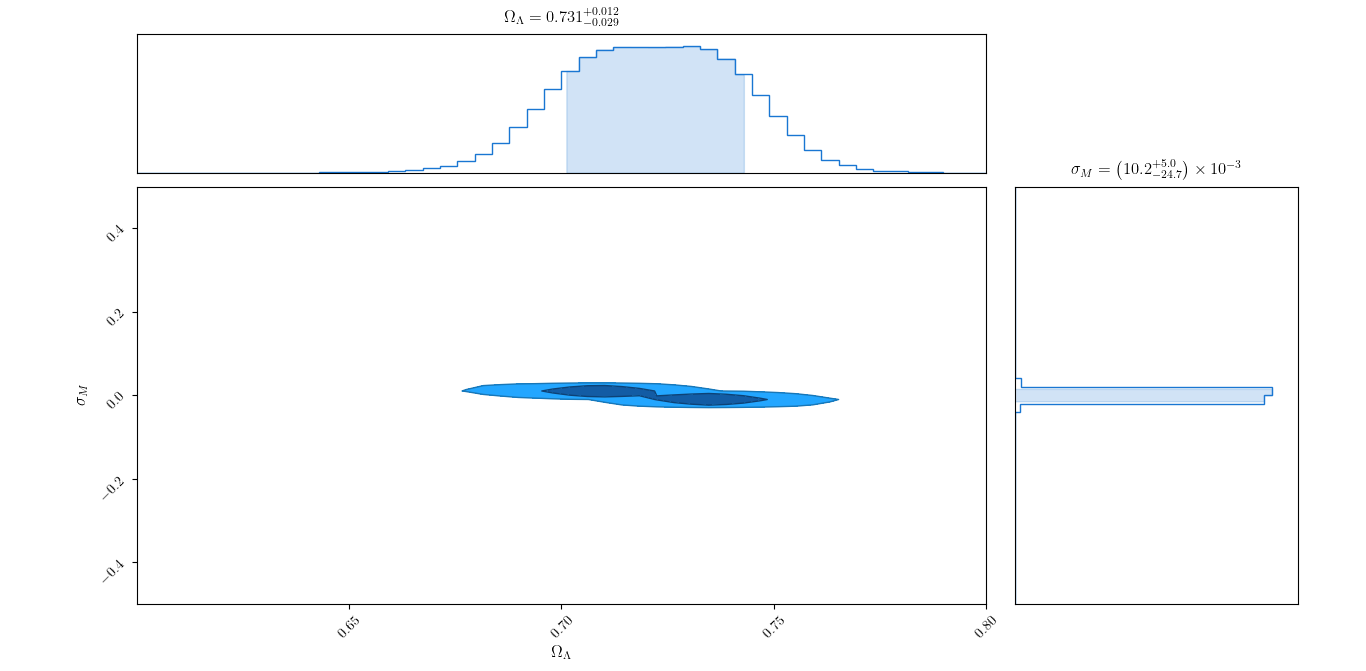}
 	\caption[Maximum likelihood estimates for Union 2.1 type Ia supernovae for $\Omega$$\Lambda$ and $\sigma_{M}$ for H0=70.]{Maximum likelihood estimates for Union 2.1 type Ia supernovae for $\Omega$$\Lambda$ and $\sigma_{M}$ for H0=70}
 	\label{fig:lambdacdmunion2-1h070sigmam-p5maxll}
 \end{figure}

\begin{figure}
 		\centering
 		\includegraphics[width=0.7\linewidth]{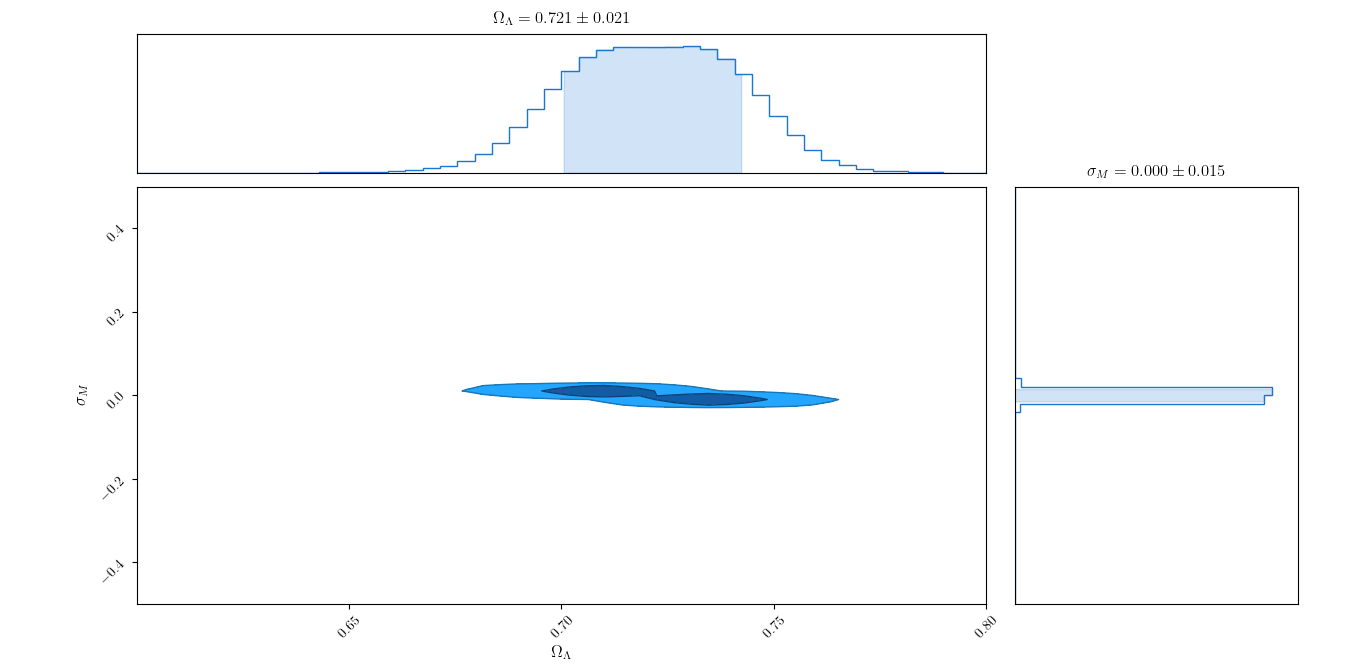}
 		\caption[Mean likelihood estimates for Union 2.1 type Ia supernovae for $\Omega$$\Lambda$ and $\sigma_{M}$ for H0=70.]{Mean likelihood estimates for Union 2.1 type Ia supernovae for $\Omega$$\Lambda$ and $\sigma_{M}$ for H0=70}
 		\label{fig:lambdacdmunion2-1h070sigmam-p5meanll}
 \end{figure}

\begin{figure}
 	\centering
 	\includegraphics[width=0.7\linewidth]{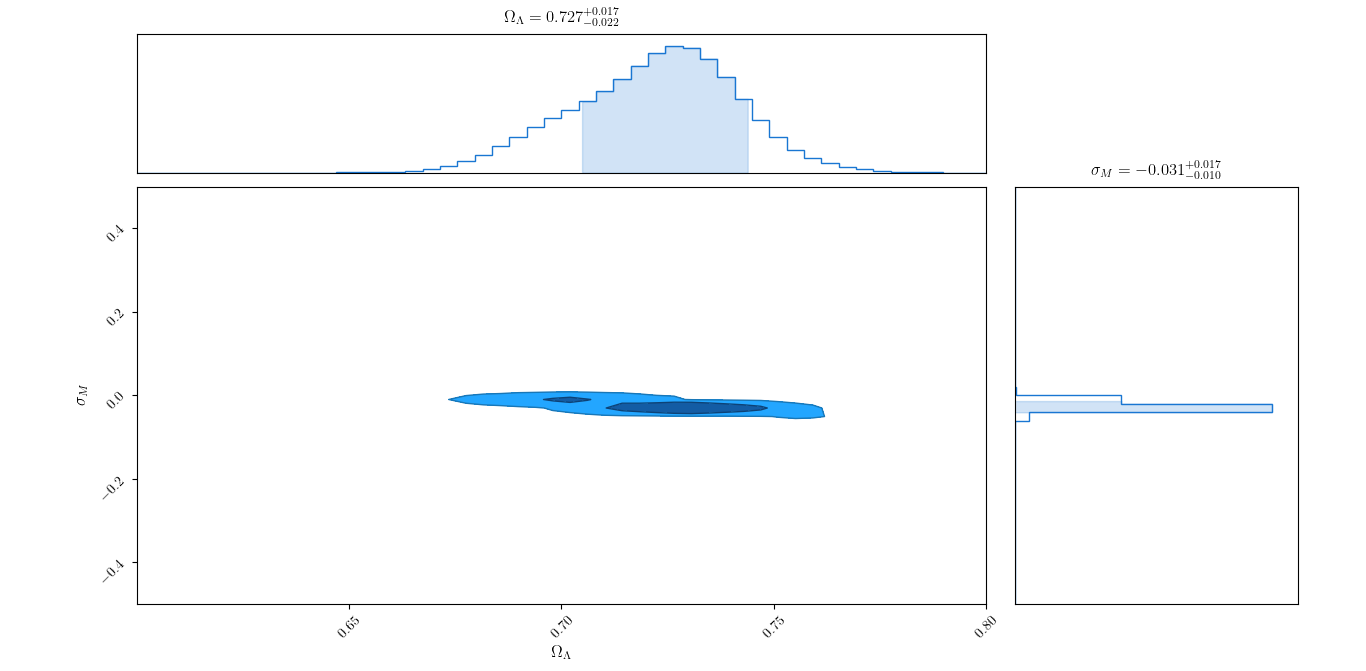}
 	\caption[Maximum likelihood estimates for Union 2.1 type Ia supernovae for $\Omega$$\Lambda$ and $\sigma_{M}$ for H0$\approxeq$69.185.]{Maximum likelihood estimates for Union 2.1 type Ia supernovae for $\Omega$$\Lambda$ and $\sigma_{M}$ for H0$\approxeq$69.185.}
 	\label{fig:lambdacdmunion2-1h069-185sigmam-p5maxll}
 \end{figure}

\begin{figure}
	\centering
	\includegraphics[width=0.7\linewidth]{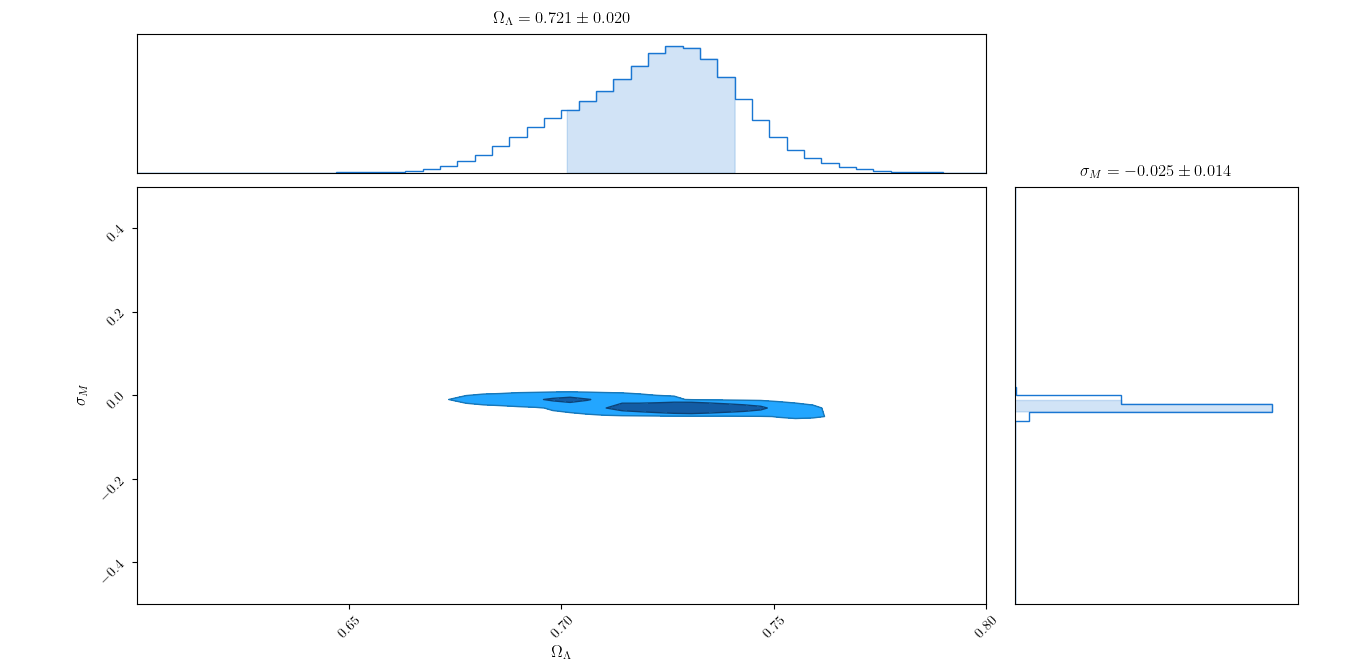}
	\caption[Mean likelihood estimates for Union 2.1 type Ia supernovae for $\Omega$$\Lambda$ and $\sigma_{M}$ for H0$\approxeq$69.185.]{Mean likelihood estimates for Union 2.1 type Ia supernovae for $\Omega$$\Lambda$ and $\sigma_{M}$ for H0$\approxeq$69.185.}
	\label{fig:lambdacdmunion2-1h069-185sigmam-p5meanll}
\end{figure}

\begin{figure}
	\centering
	\includegraphics[width=0.7\linewidth]{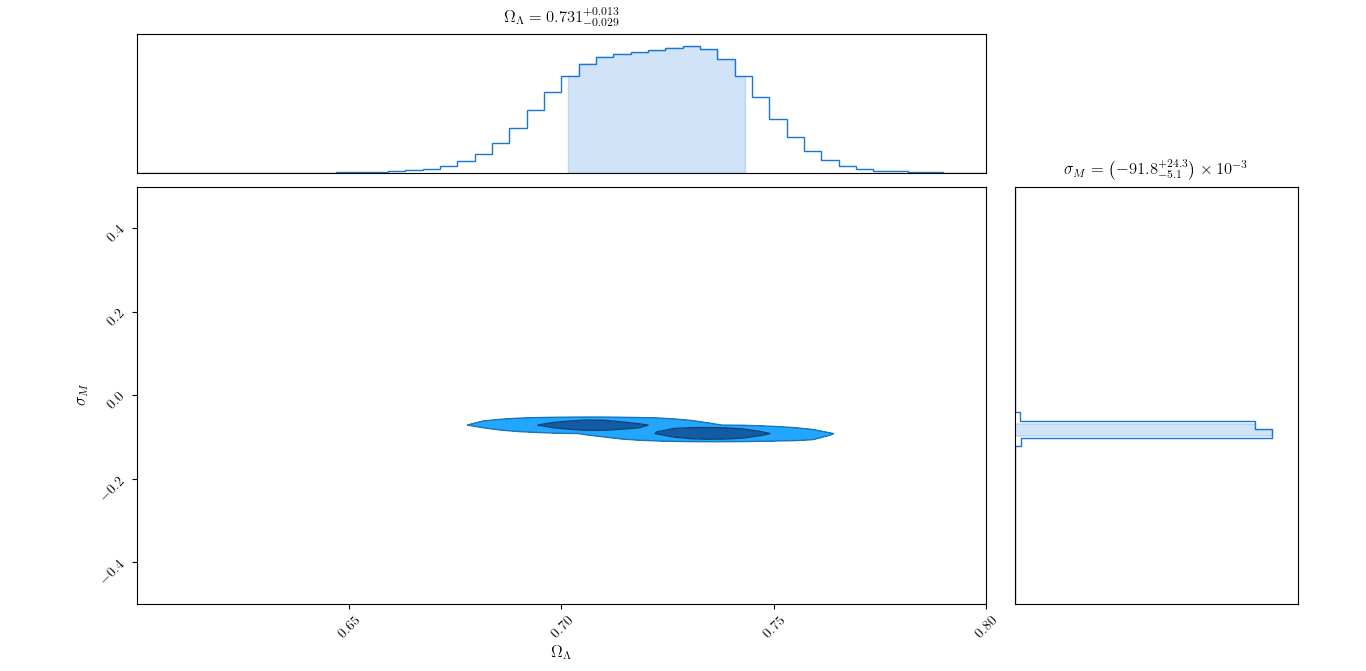}
	\caption[Maximum likelihood estimates for Union 2.1 type Ia supernovae for $\Omega$$\Lambda$ and $\sigma_{M}$ for H0=67.4.]{Maximum likelihood estimates for Union 2.1 type Ia supernovae for $\Omega$$\Lambda$ and $\sigma_{M}$ for H0=67.4.}
	\label{fig:lambdacdmunion2-1h067-4sigmam-p5maxll}
\end{figure}

\begin{figure}
	\centering
	\includegraphics[width=0.7\linewidth]{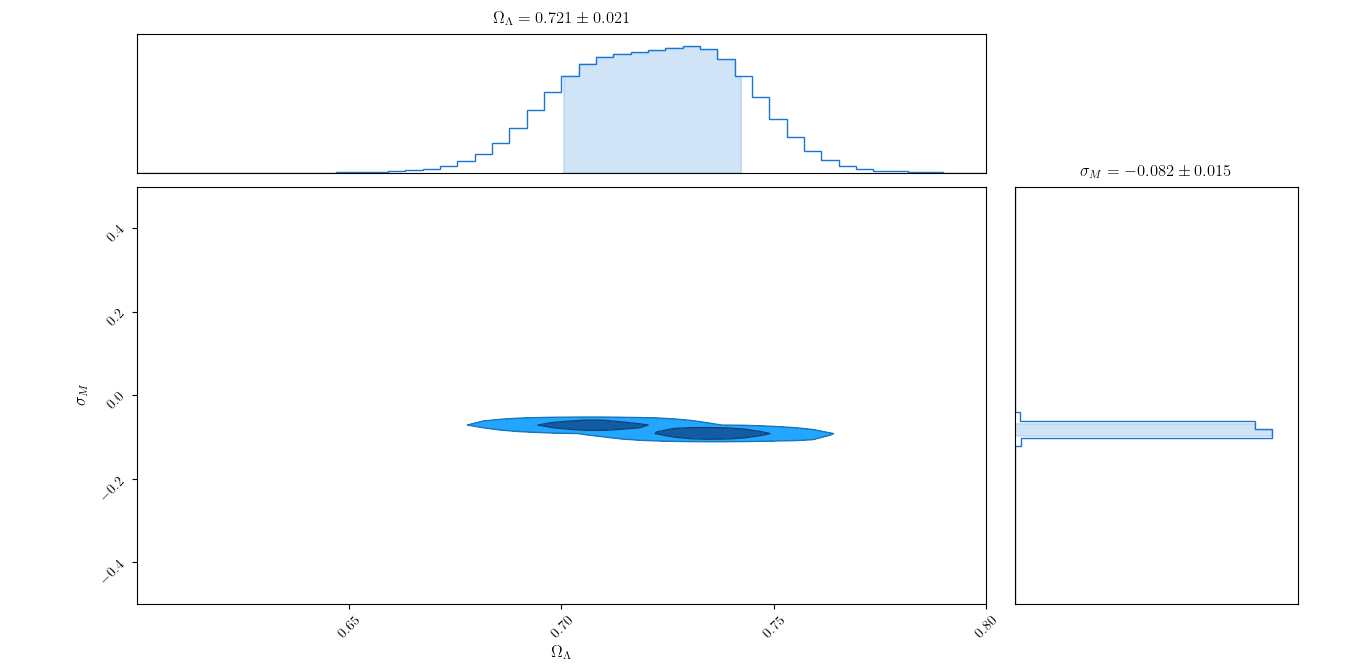}
	\caption[Mean likelihood estimates for Union 2.1 type Ia supernovae for $\Omega$$\Lambda$ and $\sigma_{M}$ for H0=67.4.]{Mean likelihood estimates for Union 2.1 type Ia supernovae for $\Omega$$\Lambda$ and $\sigma_{M}$ for H0=67.4.}
	\label{fig:lambdacdmunion2-1h067-4sigmam-p5meanll}
\end{figure}

\begin{figure}
 	\centering
 	\includegraphics[width=0.7\linewidth]{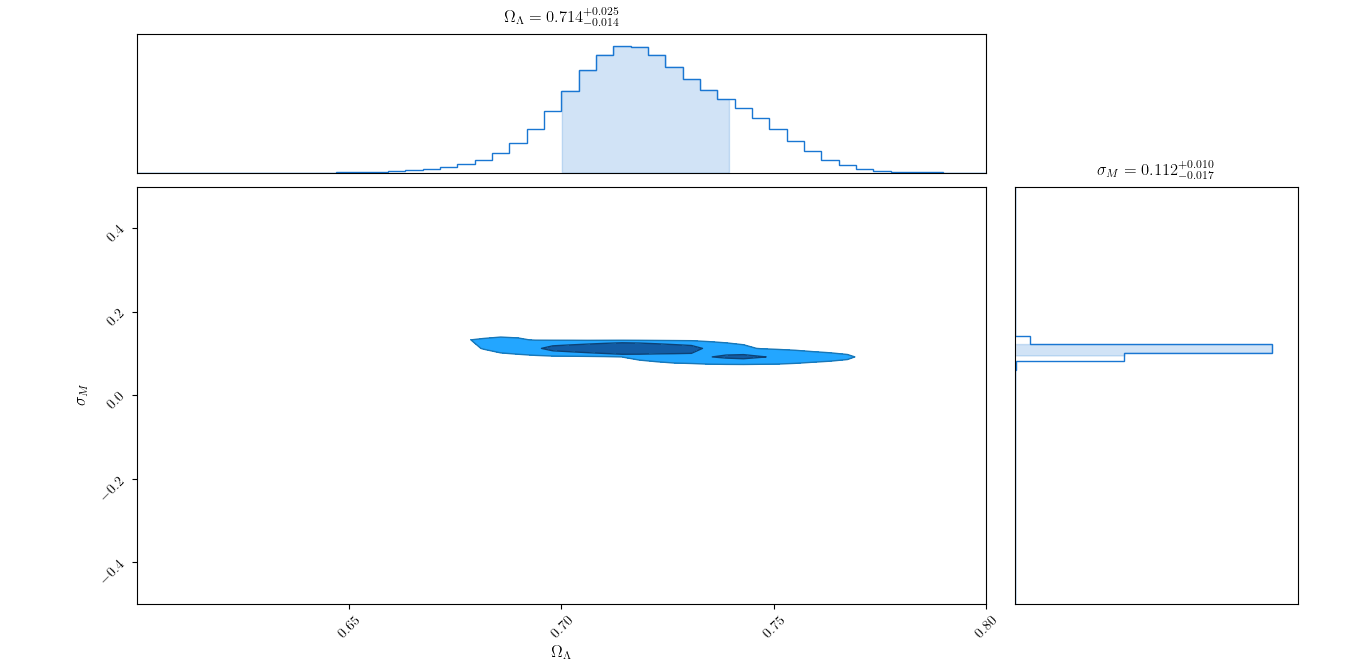}
 	\caption[Maximum likelihood estimates for Union 2.1 type Ia supernovae for $\Omega$$\Lambda$ and $\sigma_{M}$ for H0=73.52.]{Maximum likelihood estimates for Union 2.1 type Ia supernovae for $\Omega$$\Lambda$ and $\sigma_{M}$ for H0=73.52.}
 	\label{fig:lambdacdmunion2-1h073-52sigmam-p5maxll}
 \end{figure}

\begin{figure}
 	\centering
 	\includegraphics[width=0.7\linewidth]{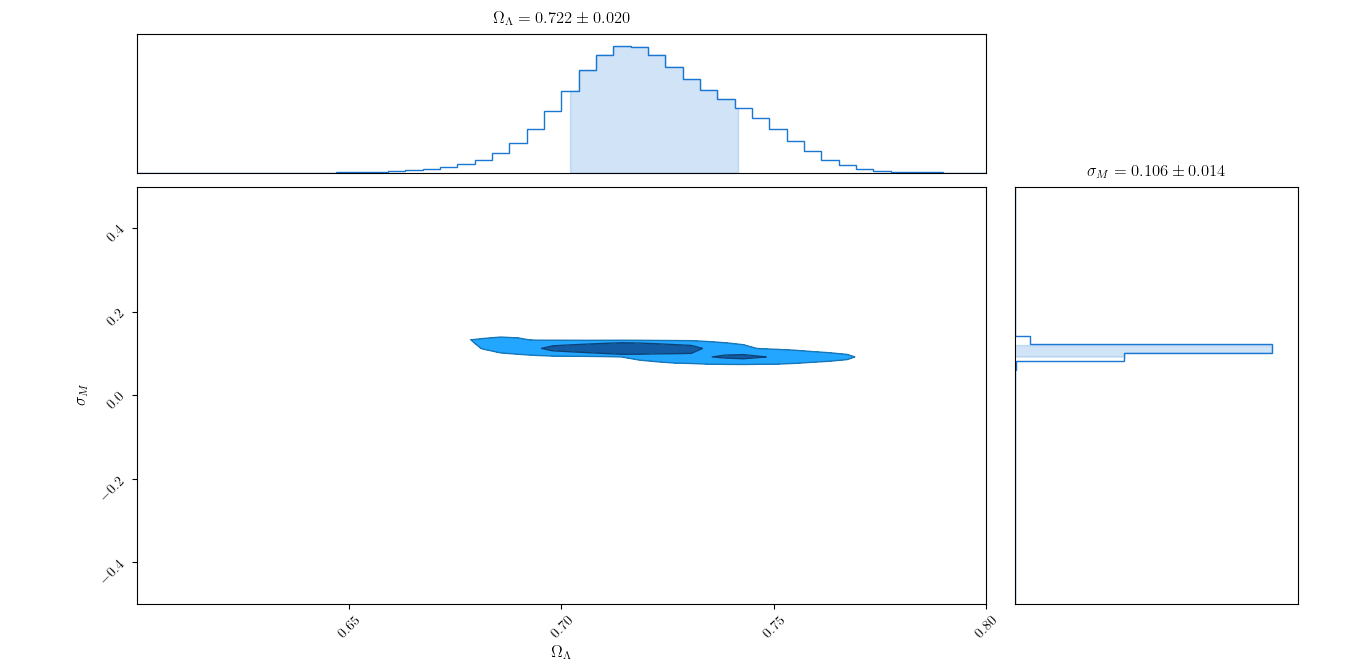}
 	\caption[Mean likelihood estimates for Union 2.1 type Ia supernovae for $\Omega$$\Lambda$ and $\sigma_{M}$ for H0=73.52.]{Mean likelihood estimates for Union 2.1 type Ia supernovae for $\Omega$$\Lambda$ and $\sigma_{M}$ for H0=73.52.}
 	\label{fig:lambdacdmunion2-1h073-52sigmam-p5meanll}
 \end{figure}

The difference between results from various H0 values can also arise from the degeneracy between the Hubble Constant value and uncertainty in absolute value (M) of type Ia supernovae which is shown in equation \ref{eq:sigma_M} as $\sigma_{M}$. However if H0 is properly fit or set then the degeneracy between H0 and M should not play much part as $\sigma_{M}$ should be more concerned about the physical properties which relate to the luminosity of the type Ia supernovae in question instead of a cosmological parameter like H0. We can see in figures \ref{fig:lambdacdmunion2-1h070sigmam-p5maxll}and \ref{fig:lambdacdmunion2-1h070sigmam-p5meanll} is $\sigma_{M}$ $\approxeq$0 for H0=70 especially in mean likelihood case which is less vulnerable to parameter boundary cuts and grid size. However, for H$\approxeq$69.185, the uncertainty increases a bit as shown in figures \ref{fig:lambdacdmunion2-1h069-185sigmam-p5maxll} and \ref{fig:lambdacdmunion2-1h069-185sigmam-p5meanll}. For H0=73.52 \cite{Riess et al. 2018b} and H0=67.4 \cite{Planck 2018}, the uncertainty increases further, as shown in figures( \ref{fig:lambdacdmunion2-1h067-4sigmam-p5maxll}, \ref{fig:lambdacdmunion2-1h067-4sigmam-p5meanll}, \ref{fig:lambdacdmunion2-1h073-52sigmam-p5maxll} and \ref{fig:lambdacdmunion2-1h073-52sigmam-p5meanll}), which may indicate that for less optimal H0 values for a given dataset, $\sigma_{M}$ starts working as an offset parameter for the models being tested instead of only representing uncertainties in the absolute value,M. In order to make more effective use of type Ia supernovae as tools to constraint cosmological parameters, we need better approaches to fit magnitude related parameters which are independent of H0 and other cosmological model considerations. However with current available data and techniques, if we apply proper fits for coefficients for stretch, color and probability of supernova in data are hosted by galaxies with less than certain threshold mass then $\sigma_{M}$ should not play much part. For this reason in tables \ref{tab:Table1},\ref{tab:Table2} and \ref{tab:Table3} we did not consider $\sigma_{M}$ in our analysis using TRF and dog leg (dogbox) methods. This will prevent $\sigma_{M}$ from working as an offset parameter and will help us in studying the relation between H0 and dark energy parameters, and so can help in at least testing various H0 values coming from different studies . 
 

 
\section{Conclusion}
We studied various dynamic dark energy EoS models and also discussed the key EoS parameter w0 in relation with the Hubble Constant. We also observed strong negative correlation between the Hubble Constant and EoS parameter w0. This relation is also studied in relation with different H0 values obtained from various surveys adopting different techniques to constraint the cosmological parameters especially H0. We found that the models we tested agreed mostly with standard cosmological model predictions. We also observed that the extended dynamic dark energy equation of state (EoS) models we tested agreed with the idea of a universe going through an accelerated expansion phase. We also observed that the value of w0, which provides value of $w_{de}$(z) at z=0 or the current epoch, is in quite close to the standard $\Lambda$-CDM constant value of $w_{de}$=-1 with w0=-1 in the confidence interval of one sigma. For the Hubble Constant value of H0 $\approx$ 69.185, which we fit on Union 2.1 dataset using kinematic expression for luminosity distance, we found that best fit values for dynamic dark energy EoS models deviate from the constant $w_{de}$=-1. However, $\Lambda$-CDM with constant value of $w_{de}$=-1 still comes as the preferable model based on the BIC selection criteria. However this deviation, even in the EoS models with higher number of parameters, shows the importance of studying H0 in relation with $w_{de}$(z). Based on our results, we can also conclude that by carefully modeling and studying $w_{de}$(z), we can potentially resolve the Hubble Constant tension arising from the results obtained using different techniques. However, to achieve this we may need to develop better methods to fit magnitude values independent of H0 and other cosmological parameters.

\section{Acknowledgement}
I would like to thank Prof. Dr. Jeremy Mould, Emeritus Professor at Swinburne University of Technology for reviewing this work and providing useful suggestions during the development of this paper.
\clearpage

\begin{table}
	
	\tiny
	\caption[short text]{ Best Fit Parameters for Dynamic Dark Energy EoS models using Union 2.1 SN type Ia dataset}
	\begin{tabular}{lllllllllll}
		\hline
		\multicolumn{1}{l}{H0} &  EoS Models &         &         & Parameter Values &         &         &         &         &         &  \\
		
		&         & Omega\_Lambda & w0      & wa      & \multicolumn{1}{l}{wb} & $a_{t}$      & T       & \multicolumn{1}{l}{Chi-Square} & \multicolumn{1}{l}{Para} & \multicolumn{1}{l|}{BIC} \\
		\hline
		\multicolumn{1}{l}{\textbf{70}} &         &         &         &         &         &         &         &         &         &  \\
		\hline
		& $\Lambda$-CDM & \textcolor[rgb]{ .133,  .133,  .133}{0.722287 ± 0.013} & \multicolumn{1}{r}{-1} &         &         &         &         & 562.2267 & 1       & 568.5897 \\
		& WCDM    & \textcolor[rgb]{ .133,  .133,  .133}{0.720362 ± 0.0626} & \textcolor[rgb]{ .133,  .133,  .133}{-1.00449 ± 0.1435} &         &         &         &         & 562.2257 & 2       & 574.9518 \\
		& CPL     & \textcolor[rgb]{ .133,  .133,  .133}{0.71933 ± 0.27885} & \textcolor[rgb]{ .133,  .133,  .133}{-1.00547 ± 0.291303} & \textcolor[rgb]{ .133,  .133,  .133}{-0.01126 ± 3.033239} &         &         &         & 562.2257 & 3       & 581.3148 \\
		& JBP     & \textcolor[rgb]{ .133,  .133,  .133}{0.70796 ± 0.17292} & \textcolor[rgb]{ .133,  .133,  .133}{-1.011969 ± 0.1643} & \textcolor[rgb]{ .133,  .133,  .133}{-0.22393 ± 3.2172} &         &         &         & 562.2219 & 3       & 581.311 \\
		& BA      & \textcolor[rgb]{ .133,  .133,  .133}{0.75 ± 0.86891} & \textcolor[rgb]{ .133,  .133,  .133}{-0.9691 ± 1.06464} & \textcolor[rgb]{ .133,  .133,  .133}{0.15139 ± 3.56007} &         &         &         & 562.2117 & 3       & 581.3008 \\
		& PADE-I  & \textcolor[rgb]{ .133,  .133,  .133}{0.7196 ± 1.24524} & \textcolor[rgb]{ .133,  .133,  .133}{-1.00523 ± 1.94515} & \textcolor[rgb]{ .133,  .133,  .133}{-0.00578414 ± 4909.4} & \multicolumn{1}{l}{\textcolor[rgb]{ .133,  .133,  .133}{-0.002752 ±  4882.5}} &         &         & 562.2257 & 4       & 587.6778 \\
		& PADE-II & \textcolor[rgb]{ .133,  .133,  .133}{0.712876 ± 0.4852} & \textcolor[rgb]{ .133,  .133,  .133}{-1.01089 ± 0.7675} & \textcolor[rgb]{ .133,  .133,  .133}{1.10857 ± 271.43} & \multicolumn{1}{l}{\textcolor[rgb]{ .133,  .133,  .133}{-0.998715 ± 267.121}} &         &         & 562.2249 & 4       & 587.6771 \\
		& LH4     & \textcolor[rgb]{ .133,  .133,  .133}{0.72007 ± 2.0615} & \textcolor[rgb]{ .133,  .133,  .133}{-0.999102 ± 223.47} & \textcolor[rgb]{ .133,  .133,  .133}{-1.01078 ± 314.52} &         & \textcolor[rgb]{ .133,  .133,  .133}{0.9379244 ± 29648.86} & \textcolor[rgb]{ .133,  .133,  .133}{0.960383 ± 39515.18} & 562.2258 & 5       & 594.0409 \\
		\hline
		\multicolumn{1}{l}{\textbf{69.185}} &         &         &         &         &         &         &         &         &         &  \\
		\hline
		& $\Lambda$-CDM & \textcolor[rgb]{ .133,  .133,  .133}{ 0.6867 ± 0.013664} & \multicolumn{1}{r}{-1} &         &         &         &         & 567.9516 & 1       & 574.3146 \\
		& WCDM    & \textcolor[rgb]{ .133,  .133,  .133}{0.75 ± 0.08088} & \textcolor[rgb]{ .133,  .133,  .133}{-0.86452 ± 0.1448} &         &         &         &         & 565.9402 & 2       & 578.6663 \\
		& CPL     & \textcolor[rgb]{ .133,  .133,  .133}{0.65 ± 0.094414} & \textcolor[rgb]{ .133,  .133,  .133}{-0.82971 ± 0.112769} & \textcolor[rgb]{ .133,  .133,  .133}{-2.27778±2.70177} &         &         &         & 564.2394 & 3       & 583.3285 \\
		& JBP     & \textcolor[rgb]{ .133,  .133,  .133}{0.65 ± 0.08967} & \textcolor[rgb]{ .133,  .133,  .133}{-0.77572± 0.13167} & \textcolor[rgb]{ .133,  .133,  .133}{-3.38651 ± 3.53197} &         &         &         & 563.8489 & 3       & 582.938 \\
		& BA      & \textcolor[rgb]{ .133,  .133,  .133}{0.65 ± 0.10221} & \textcolor[rgb]{ .133,  .133,  .133}{-0.88266 ± 0.1076} & \textcolor[rgb]{ .133,  .133,  .133}{-1.260334 ± 1.7388} &         &         &         & 564.707 & 3       & 583.796 \\
		& PADE-I  & \textcolor[rgb]{ .133,  .133,  .133}{0.66386 ± 0.12523} & \textcolor[rgb]{ .133,  .133,  .133}{-0.88078055 ± 0.11613} & \textcolor[rgb]{ .133,  .133,  .133}{-0.299997 ± 2.02684} & \multicolumn{1}{l}{\textcolor[rgb]{ .133,  .133,  .133}{-0.9958 ± 0.0000003}} &         &         & 564.1249 & 4       & 589.577 \\
		& PADE-II & \textcolor[rgb]{ .133,  .133,  .133}{0.65 ± 0.2274} & \textcolor[rgb]{ .133,  .133,  .133}{-0.81318 ± 0.5637} & \textcolor[rgb]{ .133,  .133,  .133}{3.446955 ± 23.726} & \multicolumn{1}{l}{\textcolor[rgb]{ .133,  .133,  .133}{-0.999 ± 20.04}} &         &         & 564.1126 & 4       & 589.5647 \\
		& LH4     & \textcolor[rgb]{ .133,  .133,  .133}{0.65596 ± 0.05682} & \textcolor[rgb]{ .133,  .133,  .133}{ -0.38368 ± 0.71809} & \textcolor[rgb]{ .133,  .133,  .133}{-1.23068 ± 0.28266} &         & \textcolor[rgb]{ .133,  .133,  .133}{0.96887 ± 0.03171} & \textcolor[rgb]{ .133,  .133,  .133}{0.000875 ± 0.26241} & 561.3035 & 5       & 593.1186 \\
	\end{tabular}%
	\label{tab:Table3}%
\end{table}

\
\end{document}